\pdfoutput=1

 \documentclass{elsart}
\usepackage{epsfig}
\usepackage{epstopdf}

\usepackage{graphics,epsfig}
\usepackage{graphicx}
\usepackage{bm}
\usepackage{amsmath}
\usepackage{mathrsfs}

\newcommand{\be}{\begin{equation}}
\newcommand{\ee}{\end{equation}}
\newcommand{\beq}{\begin{eqnarray}}
\newcommand{\eeq}{\end{eqnarray}}

\def\nue{\mathrel{{\nu_e}}}
\def\numu{\mathrel{{\nu_\mu}}}
\def\nutau{\mathrel{{\nu_\tau}}}
\def\nux{\mathrel{{\nu_x}}}

\def\barnue{\mathrel{{\bar \nu}_e}}
\def\barnumu{\mathrel{{\bar \nu}_\mu}}
\def\barnutau{\mathrel{{\bar \nu}_\tau}}
\def\barnux{\mathrel{{\bar \nu}_x}}

\def \gta {\mathrel{\vcenter{\hbox{$>$}\nointerlineskip\hbox{$\sim$}}}}

\def\t13{\mathrel{{\theta_{13}}}}
\def\y12{\mathrel{{\tan^2 \theta_{12}}}}
\def\c2{\mathrel{{\chi^2 }}}

\newcommand{\df}{DSN$\nu$F}

\newcommand{\sk}{SK}
\newcommand{\ceci}[1]{{}}

\newcommand{\n}{neutrino}
\newcommand{\ns}{neutrinos}

\newcommand{\sn}{supernova}
\newcommand{\sne}{supernovae}

\begin{document}
\begin{frontmatter}
\hbox to\hsize{\hfill RBRC 738}

\title{Upper limits on the diffuse supernova neutrino flux from the SuperKamiokande data}

\author{Cecilia Lunardini}
\address{Arizona State University, Tempe, AZ 85287-1504}%
\address{RIKEN BNL Research Center, Brookhaven National Laboratory, Upton, NY 11973}
\ead{ Cecilia.Lunardini@asu.edu}

\author{Orlando L. G. Peres}
\address{Instituto de F\'\i sica Gleb Wataghin - UNICAMP, 
  13083-970 Campinas SP, Brazil}
\ead{orlando@ifi.unicamp.br}

\begin{abstract}
We analyze the 1496 days of SuperKamiokande data to put limits on the
$\nue$, $\barnue$, $\numu+\nutau$ and $\barnumu+\barnutau$ components
of the diffuse flux of supernova neutrinos, in different energy
intervals and for different neutrino energy spectra.  By considering
the presence of only one component at a time, we find the following
bounds at 90\% C.L. and for neutrino energy $E>19.3$ MeV: $\Phi_{\nue}
<73.3-154~{\rm cm^{-2} s^{-1}}$, $\Phi_{\barnue} <1.4-1.9~{\rm
cm^{-2} s^{-1}}$,
$\Phi_{\numu+\nutau} <(1.0-1.4)\cdot 10^3~{\rm cm^{-2} s^{-1}}$ and
$\Phi_{\barnumu+\barnutau} <(1.3-1.8)\cdot 10^3~{\rm cm^{-2} s^{-1}}$, where the
intervals account for varying the neutrino spectrum.  
 In the interval $E = 22.9 - 36.9$ MeV, we find  $\Phi_{\nue}
<39-54~{\rm cm^{-2} s^{-1}}$, which improves on the existing limit
from SNO in the same energy window.  Our results for $\numu+\nutau$
and $\barnumu+\barnutau$ improve by about four orders of magnitude
over the previous best constraints from LSD. 
\end{abstract}


\end{frontmatter}

\section{Introduction}

The search for the diffuse flux of neutrinos from all supernovae
represents a new frontier of neutrino astrophysics.  Just like
ultra-high energy neutrinos, the diffuse supernova neutrinos
originate, for a good part, at cosmological distances. Therefore, they
could give unique insights into the history of the universe, and
specifically into the cosmological evolutions of the supernova rate
and of the star formation rate, the physics of the first stars, etc..
This possibility, to do cosmology with neutrino data, is new.  In
addition to this, the diffuse flux will be a new test of the physics of
core collapse supernovae and of neutrino propagation inside them,
which will be especially important if a galactic supernova, a very
rare event, does not occur in the next ten years or so.

Until now, the searches for the diffuse supernova neutrino flux (\df)
have turned out negative, and upper limits were put.  The strongest
limit comes from the dominant detection mode -- inverse beta decay,
$\barnue + p \rightarrow n + e^+$ -- in the largest water Cerenkov
detector available, SuperKamiokande (SK), above the threshold of $19.3$ MeV
of neutrino energy \cite{Malek:2002ns}. The limit on the $\barnue$ component of the flux
 reads:
\be
\Phi_{\barnue}(E>19.3~ {\rm MeV})<1.2~{\rm cm^{-2} s^{-1}~~~~at~ 90\% C.L.}~,
\label{sklim}
\ee
 and is valid for certain energy spectra of the neutrino flux.
This limit approaches the theoretical predictions  \cite{Bisnovatyi-Kogan:1982rd,Krauss:1983zn,woosleyetal,Totani:1995rg,Totani:1995dw,Malaney:1996ar,Hartmann:1997qe,Kaplinghat:1999xi,Ando:2002ky,Strigari:2003ig,Ando:2004sb,Ando:2004hc,Iocco:2004wd,Lunardini:2005jf,Daigne:2005xi}, thus arising the expectation that the
\df\ may be seen in the near future, and triggering several theoretical studies on
the subject (see e.g. the review \cite{Lunardini:2006em} and references therein).

Bounds on other neutrino species in the \df\ are  weaker
than the $\barnue$ limit (\ref{sklim}), due to the fact that inverse beta decay
largely dominates over other detection processes in water, and that
non-water detectors have smaller volumes than SK.  

Some attention has been devoted to  the $\nue$ component of the \df. 
The strongest bound on this
 is indirect  \cite{Lunardini:2006sn}:
\be
\Phi_{\nue}(E>19.3~ {\rm MeV})<5.5~{\rm cm^{-2} s^{-1}~~~~at~ \sim 98\% C.L.}~.
\label{cecilimit}
\ee It is found by converting the SK limit on $\barnue$,
eq. (\ref{sklim}), into a result for $\nue$, using neutrino
oscillations and the strong similarity between the fluxes of muon and
tau neutrinos and antineutrinos produced inside a supernova.

Among the direct limits on $\nue$, the strongest is 
  from the search for charged current (CC) scattering on deuterium ($\nue +
  d \rightarrow p + p + e^-$) at the Sudbury Neutrino Observatory (SNO), in the interval $22.9 - 36.9$
  MeV of neutrino energy \cite{Aharmim:2006wq}. The result is: 
\be
\Phi_{\nue}(22.9< E/{\rm MeV}< 36.9)<61-93~{\rm cm^{-2} s^{-1}~~~~at~  90\% C.L.}~,
\label{snolim}
\ee 
depending on the neutrino energy spectrum.   

The old bounds from LSD \cite{Aglietta:1992yk} are still the best for
the non-electron components of the flux, $\numu+\nutau$ ($\nux$ from
now on) and $\barnumu+\barnutau$ ($\barnux$ from now on). They are
obtained from searches of neutrino scattering on $^{12}$C, and read:
\beq
&&\Phi_{\nux}(20< E/{\rm MeV}< 100)<3 \cdot 10^7~{\rm cm^{-2} s^{-1}~~~~at~  90\% C.L.}~,\nonumber \\
&&\Phi_{\barnux}(20< E/{\rm MeV}< 100)<3.3 \cdot 10^7~{\rm cm^{-2} s^{-1}~~~~at~  90\% C.L.}~.
\label{lsdmutau}
\eeq

In this paper we examine the SK data published in \cite{Malek:2002ns} and use them to 
put limits on all the distinct components of the \df: $\nue$, $\barnue$, $\nux$ and $\barnux$.
 
For $\barnue$, we elaborate on the result (\ref{sklim}), by generalizing the SK analysis to 
a wider range of neutrino spectra,
motivated by the recent progress in the theory. Our result is that the
bound (\ref{sklim}) could change by up to $\sim$50\% depending on the neutrino spectrum. 

For the $\nue$, $\nux$ and $\barnux$ components of the flux, 
the idea
is to look for events due to CC scattering of $\nue$ on oxygen, or due to
elastic scattering on electrons of neutrinos and antineutrinos of any flavor, on top of the irremovable
background.  These events are indistinguishable from inverse beta
decay, since they have the same signature in the detector, the
Cerenkov light produced by the outgoing lepton. 
One expects that, with respect to the SNO 
search for $\nue$, the larger SK volume could overcompensate the disadvantage
of the smaller detection cross section, resulting in a competitive limit.
A similar argument makes us expect a strong improvement on the LSD limits for the non-electron species, Eq. (\ref{lsdmutau}).
Our numerical analysis confirms these intuitions.

The paper is structured as follows. In sec. \ref{generalities} we give
generalities on the \df\, and discuss its energy spectrum.
Sec. \ref{dataanalysis} is a review of the SK data and of our analysis of them.
In sec. \ref{results} we present  results. Summary and discussion follow in sec. 
\ref{discussion}.

\section{Generalities}
\label{generalities}

\subsection{Neutrinos from core collapse and the diffuse flux}
\label{neutrinos}

A core collapse supernova is an extremely powerful neutrino source,
releasing about $3 \cdot 10^{53}$ ergs of energy within $\sim$10
seconds in neutrinos and antineutrinos of all flavors in similar amounts. The energy
spectrum of these neutrinos at the production point in the star is
roughly thermal, and can be described by a power law times an
exponential \cite{Keil:2002in}.  The neutrinos have average energies
in the range of 10 -- 20 MeV, with the muon and tau species having
harder spectrum than the electron ones due to their weaker (neutral
current only) coupling to matter.  

Inside the star, the neutrinos and antineutrinos undergo either
partial or total flavor conversion depending on the mixing angle
$\theta_{13}$ and on the mass hierarchy  of the neutrino
mass spectrum. The hierarchy is defined as  {\it  normal} ({\it
inverted}) if the third mass eigenstate -- the one whose electron
component is $\sin^2 \theta_{13}$ -- is the heaviest (lightest).  The
net conversion probability is due to a combination of effects of
neutrino-neutrino coherent scattering
\cite{Duan:2005cp,Duan:2006an,Hannestad:2006nj,Duan:2007mv,Raffelt:2007cb,Duan:2007fw,Duan:2007bt,Fogli:2007bk,Raffelt:2007xt,Duan:2007sh,EstebanPretel:2007ec,EstebanPretel:2007yq,Dasgupta:2007ws,Dasgupta:2008cd,Chakraborty:2008zp}
and of matter-driven resonant conversion (see e.g.,
\cite{Mikheev:1986if,Dighe:1999bi,Lunardini:2003eh}).   The neutrino-neutrino effects swap the energy
spectra of the electron and non-electron 
components of the flux for the inverted mass hierarchy and above
 a certain critical energy, $E_c$.
Typical critical energies are below $\sim$10 MeV (see
e.g. \cite{Fogli:2007bk}), with lower values for antineutrinos
compared to neutrinos.   The matter-driven conversion occurs more externally in the star and induces a further
permutation of the energy spectra of the fluxes in the different
flavors that enter the resonances.

The diffuse flux of \sn\ \ns\  in a detector is simply the sum of the
contributions from the individual stars.  This sum can be expressed as
an integral involving the cosmological rate of supernovae, $R_{
SN}(z)$, which is a function of the redshift $z$ and is defined as the
number of supernovae in the unit of  comoving volume in the unit time.
Starting with the present value, $R_{SN}(0)\sim {\mathcal
O}(10^{-4})~{\rm Mpc^{-3}~yr^{-1}}$, the rate increases with $z$ and
flattens at $z>1$ (see e.g. \cite{Hopkins:2006bw}). In terms of $R_{
SN}$,  the \df\ in a detector at Earth, differential in  energy,
surface and time, is given by:
 \be
\Phi(E)=\frac{c}{H_0}\int_0^{z_{ max}} R_{ SN}(z)\frac{{d}
N^{det} (E^\prime)}{{d} E^\prime} \frac{{d}
z}{\sqrt{\Omega_{ m}(1+z)^3+\Omega_\Lambda}}~ 
  \label{flux}
\ee
 (see
 e.g. \cite{Ando:2004hc}), where ${d}
N^{det} (E^\prime)/{d} E^\prime$ is the contribution of an individual
supernova, inclusive of neutrino oscillations and of the redshift of
energy, $E' = E(1+z)$.   $\Omega_{ m}$ and $\Omega_\Lambda$ are the fractions of the
cosmic energy density in matter and dark energy respectively;  $c$ is
the speed of light and $H_0$ is the Hubble constant.

\begin{figure}[htbp]
  \centering
\includegraphics[width=0.75\textwidth]{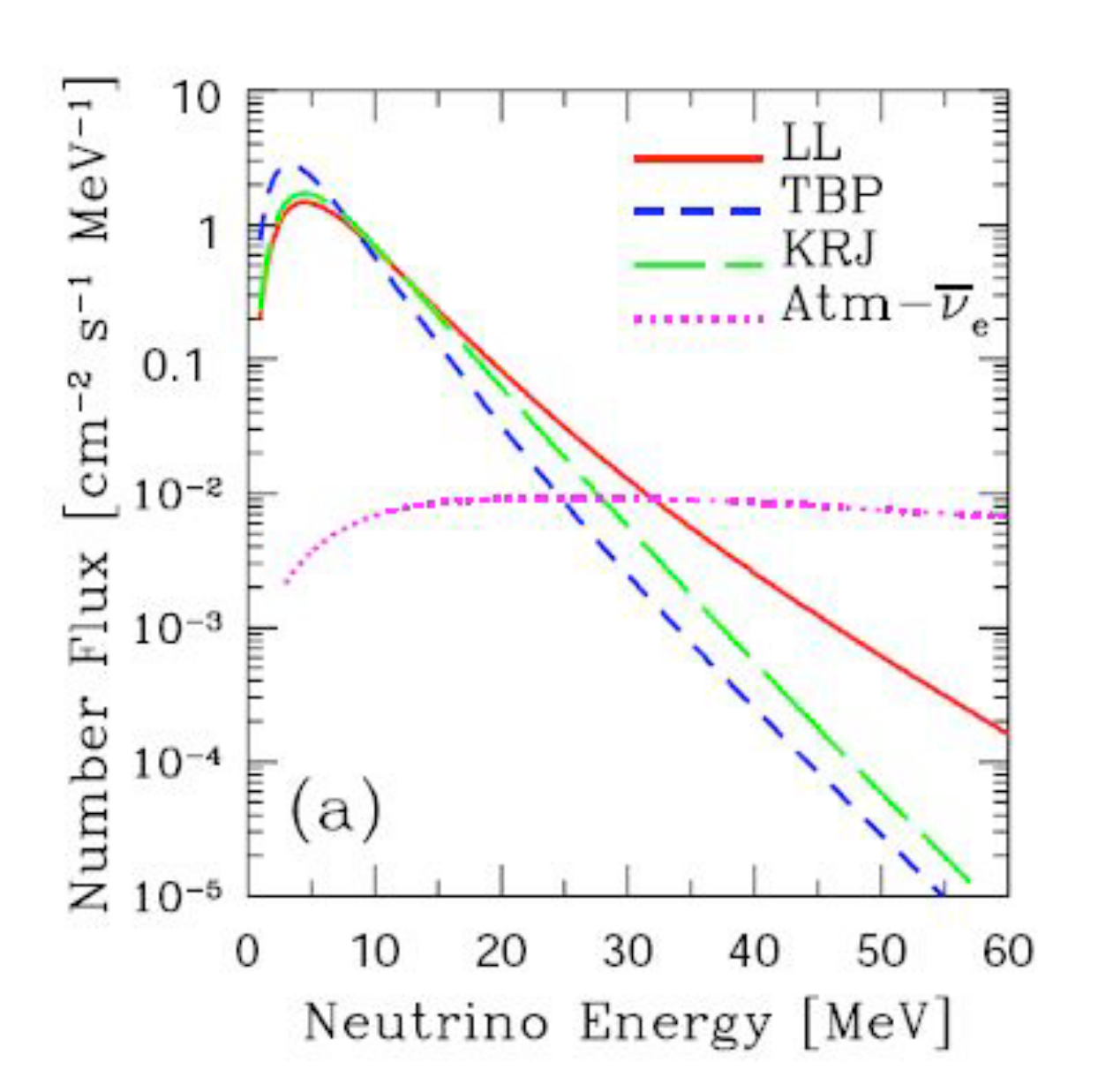}
\caption{Examples of energy spectra of the $\barnue$ component of the
\df\ from the literature, specifically from the Garching (KRJ),
Lawrence Livermore (LL) and Arizona (TBP) models,
 with oscillations (see also Table \ref{tableslopes})
\cite{Keil:2002in,Totani:1997vj,Thompson:2002mw,Ando:2004hc}. The
background due to atmospheric $\barnue$ is shown as well.   This
figure is taken from ref. \cite{Ando:2004hc}, with permission. }
\label{spectraando}
\end{figure}
Examples of calculated energy spectra for the $\barnue$ component of the \df\  are shown in fig. \ref{spectraando}. We see that the
  flux peaks at $E_{max}=3-7$ MeV, where it reaches  typical values of ${\mathcal O}(1)~{\rm cm^{-2} s^{-1} MeV^{-1}} $, and
falls rapidly at $E\gg E_{max}$.
It has been shown \cite{Lunardini:2006pd} that in this high energy regime, which is relevant for our analysis, 
the flux is
approximated very well ($\sim$5\% accuracy or better)
by an exponential: 
\be \phi(E) \simeq \phi_0
\e^{-E/E_0}~,
\label{exp}
\ee
where the energy $E_0$ is close to the energy at the peak of the
spectrum, $E_0 \sim E_{max}$.   The total flux above a realistic
detection threshold of 10-20 MeV is larger for larger $E_0$, and is
proportional to the normalization of the supernova rate, $R_{SN}$. The
large uncertainties on these two quantities translates into a large
uncertainty on the \df.

In Table \ref{tableslopes} we illustrate this by summarizing 
 several published predictions of the $\barnue$
 component of the \df.  The Table gives the values of $E_0$ found by
fitting the energy spectra with the simple exponential (\ref{exp}),
and the predicted $\barnue$ flux above the SK threshold. In both
columns we see large differences, due to the different input
 quantities and constraints that are used in calculating the \df.  
For example, to describe the \df\ energy spectrum some authors use neutrino spectra that result from
 calculations of neutrino transport (e.g., \cite{Keil:2002in}), while
 others use spectra that fit the SN1987A data 
 \cite{Lunardini:2005jf} or simply adopt thermal spectra with
 temperatures varying within indicative intervals
 \cite{Beacom:2005it}.  
This diversity of approaches  results in the variation $E_0 = 3.8 -
8.5$ MeV, which we use in our analysis\footnote{Even though the
interval of $E_0$ given here  refers to electron antineutrinos, we use
it  for other neutrino species as well, as a representative range, and
for easier comparison between different channels. }.
An even larger variation is seen in the integrated flux, which ranges
from $\sim 0.05 ~{\rm cm^{-2} s^{-1}}$ to values exceeding the SK
limit, eq. (\ref{sklim}).  

\begin{table*}
\centering
\begin{tabular}{| l | l | l |}
\hline
\hline
 model, reference and comments & $E_0/{\rm MeV}$ & $\Phi_{\barnue}(E>19.3 {\rm MeV})$  \\
   &  &  $({\rm cm^{-2} s^{-1}})$ \\
\hline
\hline 
LMA oscillations  
\cite{Ando:2002ky} & 5.68 & 0.43 \\
\hline
Galaxy evolution 
\cite{Totani:1995dw} & 5.35 & 0.41 \\
\hline
Constant SN rate  
\cite{Totani:1995rg} & 5.62 & 3.1 \\
\hline 
Cosmic gas infall \cite{Malaney:1996ar}   & 5.3 & 0.2 \\
\hline 
Cosmic chemical evolution  \cite{Hartmann:1997qe}  (``NC'' model)  & 5.1 & 0.39 \\
\hline 
Heavy metal abundance \cite{Kaplinghat:1999xi}  & 5.1 & $<2.2$ \\
\hline
\hline
SN1987A fit \cite{Lunardini:2005jf} & 4 - 7 & 0.05-0.35 (99\% C.L.)\\
\hline
SN1987A-inspired \cite{Fukugita:2002qw}  & 3.8 - 5.8 & $\sim 0.24 - 1.2$\\
 (fig. 4 in \cite{Fukugita:2002qw}) &  & \\
\hline
``concordance'' \cite{Strigari:2003ig,Beacom:2005it} & 4.0 - 8.5 & 0.3-1.2\\
\hline
Chemical evolution, bimodal IMF \cite{Daigne:2005xi} & not given & 0.4 - 3.2\\
\hline
Garching code \cite{Keil:2002in}, & 4.4 & 0.28\\
 with LMA oscillations \cite{Ando:2004hc}  (fig. \ref{spectraando}) &  &  \\
\hline
Lawerence Livermore code \cite{Totani:1997vj}, & 5.2 & 0.46\\
 with LMA oscillations \cite{Ando:2004hc}  (fig. \ref{spectraando}) &  &  \\
\hline
Arizona code \cite{Thompson:2002mw}, & 3.9 & 0.14\\
 with LMA oscillations \cite{Ando:2004hc}  (fig. \ref{spectraando}) &  &  \\
\hline
\hline
\end{tabular}
\caption{Summary of existing calculations of the $\barnue$ component
of the \df.  For each, we give the value of $E_0$ for which a simple
exponential spectrum, Eq.  (\ref{exp}), best fits  the \n\ spectrum
in the high energy regime ($E \gta 10$ MeV).  We also give the
integrated flux above the SK energy threshold.  
The models in the first six rows are those considered in the SK
analysis \cite{Malek:2002ns}.  In several cases, the values quoted in
the table have been inferred from graphics in the original
references. The fluxes in ref. \cite{Ando:2004hc} are calculated up to
a normalization factor, which is estimated to be of order unity. }
\label{tableslopes}
\end{table*}

In closing this section, we would like to point out that the effects
of neutrino-neutrino scattering on flavor conversion have been studied
in detail only recently, and therefore they were not included in the
literature we have referenced for the spectral shape, eq. (\ref{exp}),
and for the results in figure \ref{spectraando} and in Table
\ref{tableslopes}.  While the inclusion of these effects is certainly
in the agenda for the near future (see the initial study in \cite{Chakraborty:2008zp}), we believe that the literature we
have quoted is still a good description for the purpose of this paper.
Indeed, older results are still valid for the normal mass hierarchy,
where neutrino-neutrino effects are negligible. Moreover, they can
reproduce (with effective parameters) the case of inverted mass
hierarchy with  neutrino-neutrino scattering above the swap energy
$E_c$. This is sufficient for our analysis, since $E_c$ is typically
below the energy windows of interest here, $E \gta 10$ MeV (see Table
\ref{tabfinal}; a possible exception is LENA at Wellington, as the
table shows).

\section{The data and the analysis}
\label{dataanalysis}

\subsection{The data and their interpretation}
\label{data}
\begin{figure}[htbp]
 \centering
 \includegraphics[width=1.1\textwidth]{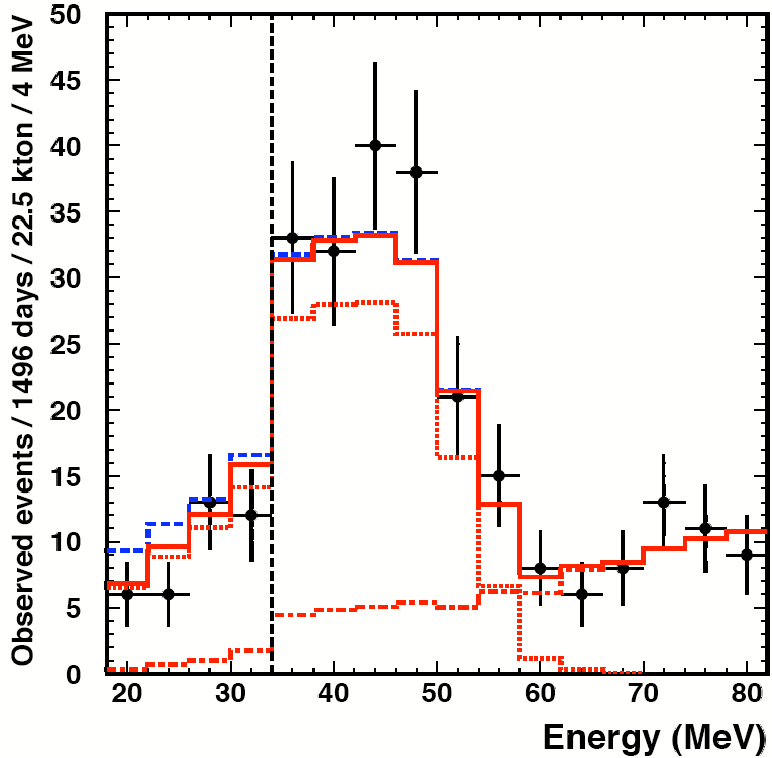}
\caption{The  distribution in lepton energy of the SK events and of the fitted
backgrounds.  The dotted and dashed histograms below the solid one are the fitted
backgrounds from invisible muons and atmospheric neutrinos ($\nue$ and $\barnue$) respectively.  The solid
histogram is the sum of these two backgrounds.  The dashed line above the solid one shows
the sum of the total background and of the 90\% upper limit on the \df\
signal found in \cite{Malek:2002ns} for the $\barnue$ component of this flux. The figure is taken from ref. \cite{Malek:2003ki}, with
permission.    Note that the figure published in \cite{Malek:2002ns}
is corrected for efficiency, i.e., it does not include the cuts that
motivate the realistic efficiency in Eq. (\ref{SKeff}). }
\label{malek_figure}
\end{figure}

We have used the electron-like events from the 1496 days of operation
of SuperKamiokande, published in ref.  \cite{Malek:2002ns}  and
described in more detail in \cite{Malek:2003ki}.  Their
energy distribution is shown in fig. \ref{malek_figure}. Due to the
cuts motivated by background rejection, the events are limited to the
interval $18-82$ MeV in lepton energy.  

The events could be due to neutrinos of different species.
Considering only the dominant interaction channel for each species, we
have the following possibilities (see also, e.g. \cite{Volpe:2007qx}):

\begin{enumerate}

\item   $\barnue$
interacting via inverse beta decay:
\beq
\barnue + p \rightarrow n + e^+ ~,
\label{invbet}
\eeq

\item   $\nue$
scattering on oxygen and on electrons:
\beq 
 \nue + ^{16}{\rm O} \rightarrow  X + e^-  \label{nueox}\\
\nue+ e^-  \rightarrow  \nue + e^- ~
\label{nuescatt}
\eeq
(here $X$ denotes all possible final states, $X=^{16}{\rm F}, ^{16}{\rm F}^\ast, ....$), 
\item   muon and tau neutrinos and antineutrinos scattering on
electrons:
\beq 
 \nux + e^- \rightarrow \nux + e^- ~, \nonumber\\
\barnux+ e^- \rightarrow \barnux + e^- ~.
\label{nuescatt1}
\eeq

\end{enumerate}
All these channels are indistinguishable in the detector because
 SK is not charge-sensitive, and so it does not discriminate between
electrons and positrons. Moreover, since the \df\ is isotropic in
space, one can not use angular information to separate elastic
scattering events,  in contrast with an individual neutrino burst.

In the natural assumption of similar luminosities in the  different
species (sec. \ref{neutrinos}), the process (\ref{invbet}) largely
dominates due to its larger cross section, hence the choice in
ref. \cite{Malek:2002ns} to consider only this channel, obtaining a
limit on the $\barnue$ component of the \df.

Here we
study all the channels above, resulting in limits on the $\barnue$, $\nue$, $\nux$ and $\barnux$ components of the \df.  We do this 
by  considering one neutrino species at a time, meaning that to derive
the limit on the $\nue$ component of the \df\ we neglected the
presence of all the other species, and likewise for the other components of the flux.
While in contrast with theory, this is an acceptable working
assumption, as it gives the most conservative upper limits.

\subsection{The analysis}
\label{analysis}

For each of the channels given in sec. \ref{data}, we performed a
$\chi^2$ analysis following the  procedure used by the SK
collaboration in \cite{Malek:2002ns,Malek:2003ki},  to which we refer
for details.   Briefly, the steps of the analysis are as follows:
\begin{enumerate}

\item We consider the events due to the \df\ (signal) and those due to
two sources of background. These are invisible muons\footnote{A muon
that has too low energy to produce any Cerenkov light in water is
called an \emph{invisible muon}. Its only observable effect is the
Cerenkov light from the electron produced by its decay. Invisible
muons originate from the interaction of atmospheric neutrinos in the
vicinity of the detector.} and  the flux of atmospheric $\nue$'s and
$\barnue$'s (figure \ref{malek_figure}).  For each of these three
components, we hold the spectral shape (taken from
\cite{Malek:2003ki}) fixed  and take the normalization as fit
parameter. Therefore, we have three parameters, one of them being the
\df\ normalization, $\phi_0$ (Eq. (\ref{exp})).

\item We calculate the $\chi^2$, which is a function of the three normalizations. It is given by the expression:
\begin{equation}
\chi^2=\sum^{16}_{l=1} \dfrac{\left(-N^{data}_l+\phi_0 A_l+\beta
B_l+\gamma C_l\right)^2}{(\sigma^{data}_l)^2+\sigma_{sys}^2}
\label{chi2}
\end{equation}
where 
$\beta$ and $\gamma$ are normalization constants for 
invisible muons and  for atmospheric 
electron neutrinos respectively.  $A_l$, $B_l$ and $C_l$ are the  number of
events from the supernova flux, invisible muons and atmospheric 
 neutrinos in the {\em l} bin. $N^{data}_l$ is the number of events in the
same bin;   $\sigma^{data}_l$ and $\sigma_{sys}$ are the statistical and systematic errors, from \cite{Malek:2002ns,Malek:2003ki}.
\item  We marginalize over the background normalizations and so obtain a function $\chi^2(\phi_0)$, which depends on the signal only.
\item From $\chi^2(\phi_0)$ we find the 90\% C.L. limit on $\phi_0$.
This limit is immediately translated into two more physically
meaningful limits: one on the \df\ and another on the number of events
due to it.  
\item  We repeat the procedure for different spectral shapes of the
diffuse supernova flux, i.e. for different values of $E_0$ in the
interval $E_0=3.8-8.5$ (Sec. \ref{generalities}).  
\end{enumerate}

To predict the histogram of events due to the \df, we have used the experimental efficiency as in \cite{Malek:2002ns,Malek:2003ki}:
\beq
 \epsilon(E_e)  = \left\{ \begin{array}{ll}
         0.47 & \mbox{if $E_e \leq 34$ MeV};\\
        0.79 & \mbox{if $E_e > 34$ MeV}.\end{array} \right.  ~,
\label{SKeff}
\eeq
and the SuperKamiokande energy resolution as given in \cite{Bahcall:1996ha}:
\be
{\mathcal E}(E_e)=0.5~ {\rm MeV} \sqrt{\frac{E_e}{{\rm MeV}}}~,
\label{SKres}
\ee
with $E_e$ being the positron/electron energy.

\begin{figure}[htbp]
  \centering
\includegraphics[width=0.90\textwidth]{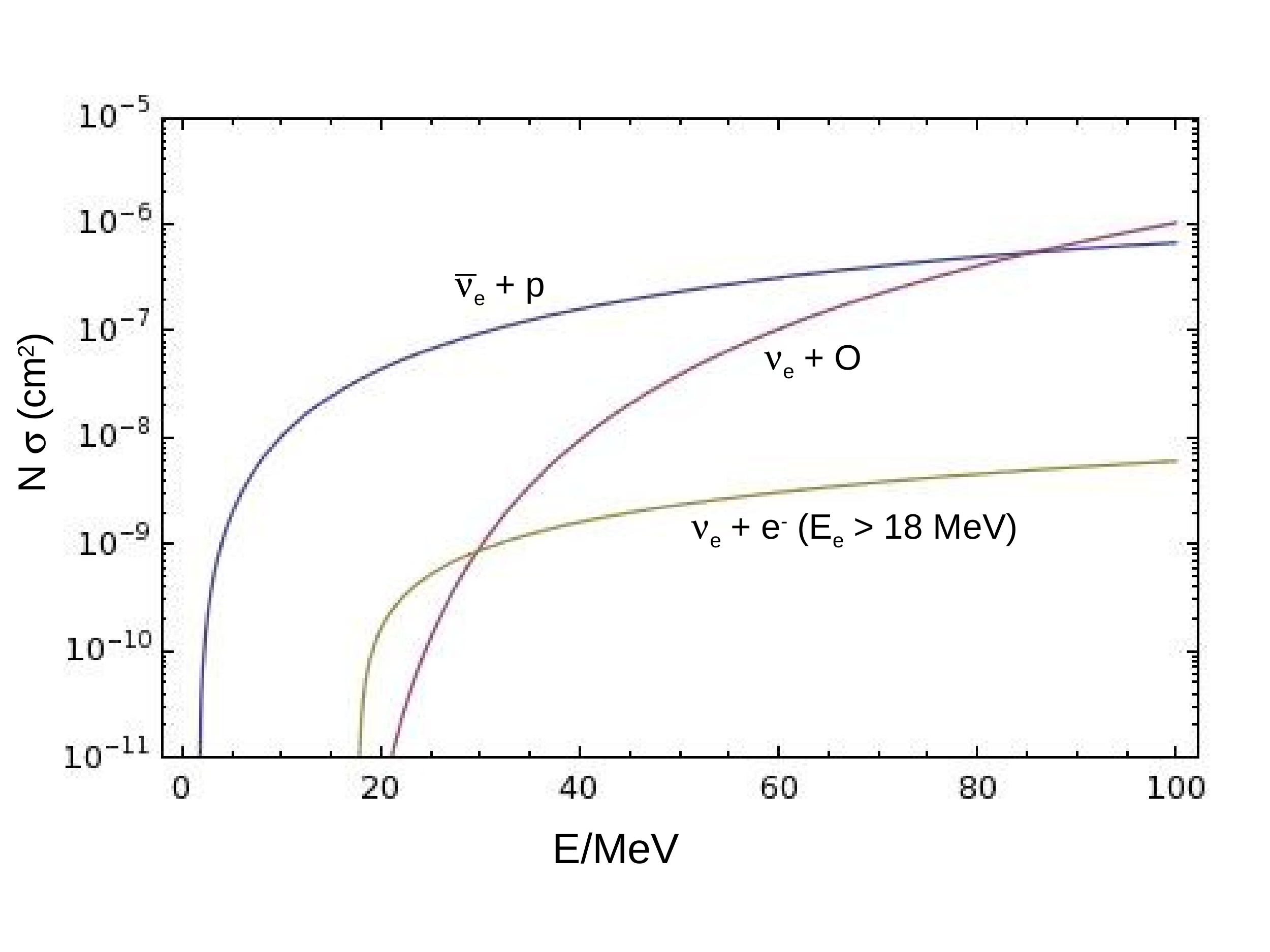}
\caption{The cross sections for the processes (\ref{invbet}),
(\ref{nueox}) and (\ref{nuescatt}), multiplied by the relevant numbers
of scatterers in SuperKamiokande, as functions of the neutrino energy,
$E$.  The cross section for $\nue$-electron scattering is integrated
over the energy of the outgoing electron, $E_e$, under the condition
that it exceeds the detector's threshold, $E_e>18$ MeV.  Note how, in
spite of the smaller cross section,  $\nue$-electron scattering
dominates over $\nue$-oxygen interaction at $E\sim 20-30$ MeV, thanks
to the larger number of scatterers.  We used the number of protons
$N_p=1.5\cdot 10^{33}$.}
\label{cross_sect}
\end{figure}

Published nuclear cross sections were adopted. In particular,  for
inverse beta decay we followed Vissani and Strumia
\cite{Strumia:2003zx} (fig. \ref{cross_sect}).  For the energy
interval considered here, nucleon recoil is negligible, therefore the
energy of the emitted positron simply differs by $1.29$ MeV from the
neutrino energy. The total cross section of $\nue$ scattering on oxygen was taken from
\cite{Kolbe:2002gk}, with the analytical form in \cite{Tomas:2003xn}
(fig. \ref{cross_sect}).  Following the latter  reference, we have
assumed that the difference between electron energy and neutrino
energy is 15 MeV in 100\% of the cases\footnote{This assumption has
only a phenomenological, approximated validity.  Improvements on this
should come from detailed  cross section calculations of neutrino
scattering on oxygen, and have not been published so far. }.   For the scattering of all species on electrons, the Standard Model
cross section was used at the lowest order in the fine structure
constant (see e.g., \cite{Tomas:2003xn}).  Note that the difference
between the energies of the incoming neutrino and of the scattered
electron is not a constant in this case.

The analysis was  restricted   to
the physical region, $\phi_0\geq 0$, by normalizing the likelihood function to 1 in
this semi-plane, as it is done in \cite{Malek:2002ns,Malek:2003ki}.   

We found that the data are not constraining enough to have $E_0$ as an additional fit parameter, instead than a fixed quantity.
That would mean having four fit variables, two for the \df\ and two
for the background. This number of degrees of freedom is too large
considering that the \df\ falls rapidly with energy, and therefore it
is constrained mainly by the twelve events in the first two energy bins
(fig. \ref{malek_figure}).

\section{Results}
\label{results}

We now present the results for each of the four channels, $\nue$, $\barnue$, $\nux$, $\barnux$. 
  They are shown in the Tables \ref{tab_barnue}-\ref{tab_barnux}, and
in the figures \ref{limits_barnue}-\ref{limits_antinumu}.  For each
channel, we give both the 90\% C.L. upper limit on the number of
events, $N_{90\%}$, as well as the corresponding bound on the neutrino
flux for certain intervals of neutrino energy. 

While 
adequate overall, the exponential spectrum we use, Eq. (\ref{exp}),  tends to
overestimate the flux at low energy (see
\cite{Lunardini:2006pd}).  Therefore our upper limits for thresholds around $10-15$ MeV are slightly more conservative
than those that could be obtained with more realistic (but
model-dependent) spectral shapes.

We do not present limits on the full fluxes (i.e., integrated over all
energies) for two reasons. The first is that below $E\sim 10$ MeV the
energy spectrum of the \df\ becomes more model dependent and therefore
any bound could be given only for very specific scenarios of \n\
spectra and luminosities, supernova rate, etc., as done in \cite{Malek:2002ns}.
The second reason is that the study of the full flux is out of the
reach of current and planned experiments, due to background at low
energy\footnote{Even liquid scintillator detectors
\cite{MarrodanUndagoitia:2006re,Wurm:2007cy}, as well as the upgraded
SK with the addition of Gadolinium \cite{Beacom:2003nk}, would be limited to a $\sim$ 10
MeV threshold, due to the ineliminable background of reactor
neutrinos. Lowering the threshold could be feasible if a
detector  becomes available in a region
free from sources of nuclear power \cite{Wurm:2007cy}.}.

\subsection{Results for $\barnue$}
\label{barnueresults}

\begin{figure}[htbp]
  \centering
\includegraphics[width=0.90\textwidth]{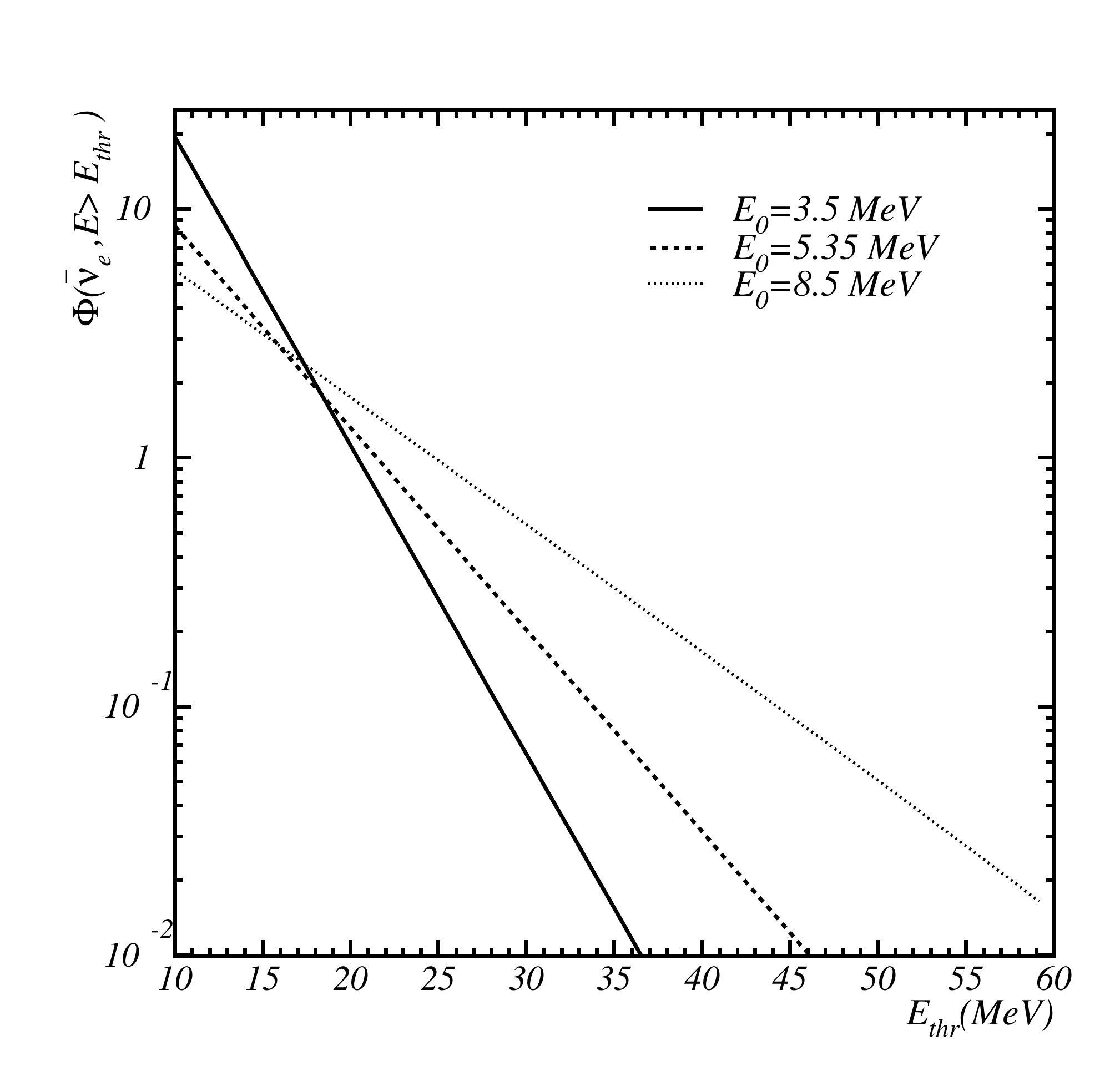}
\caption{The 90\% C.L. upper limit on the $\barnue$ flux (in ${\rm
cm^{-2} s^{-1}}$) above a neutrino energy threshold $E_{thr}$, as a
function of $E_{thr}$, for different values of $E_0$. }
\label{limits_barnue}
\end{figure}

Our results for the $\barnue$ flux are summarized in Table
\ref{tab_barnue} and fig. \ref{limits_barnue}. They generalize the
study of the SK  collaboration \cite{Malek:2002ns}, where only models with $E_0 \simeq
 5.1-5.7$ MeV were considered. 
We give the flux limits 
above two thresholds of neutrino energy: $E_{thr}=19.3$ MeV and
$E_{thr}=11.3$ MeV. The first corresponds to the energy window used in
the SK analysis in \cite{Malek:2002ns}, while the second is the
expected threshold for SK upgraded with the addition of  Gadolinium to
the water \cite{Beacom:2003nk} (see sec. \ref{discussion}).  
 \begin{table*}
\centering
\begin{tabular}{| l || l || l | l |}
\hline
\hline
$E_0/{\rm MeV}$  & $N_{90\%}$  & \multicolumn{2}{|c|}{limit on $\Phi_{\barnue}$ (${\rm cm^{-2} s^{-1}}$)} \\ 
\hline
\hline
 &  & $E/{\rm MeV}>11.3$  & $E/{\rm MeV}>19.3$ \\
\hline
\hline
3.5 & 4.89 & 13.4 & 1.37 (1.16) \\
\hline
5.35 & 6.62 & 6.88  & 1.49 (1.28) \\
\hline
5.5 & 6.86 & 6.41  & 1.52 (1.29) \\
\hline
6.5 &  8.40 & 5.56  & 1.62 (1.38) \\
\hline
7.5 & 10.19  & 5.01 & 1.75 (1.50) \\
\hline 
8.5 & 12.31 & 4.87 & 1.90 (1.63) \\
\hline
\hline
\end{tabular}
\caption{Upper  limits for the $\barnue $ channel. We give the  90\%
C.L. bound for  the number of events, $N_{90\%}$, and the
corresponding limit on the flux of neutrinos above two different
energy thresholds. The numbers in brackets in the second column were
obtained using the less accurate cross section adopted in the SK
analysis \cite{Malek:2002ns}; they are shown for comparison. }
\label{tab_barnue}
\end{table*}

The 90\% C.L. bound on the number of
events is in the range $N_{90\%}\sim 5 - 12$, and the corresponding 
 upper limit on the flux above 19.3 MeV of neutrino energy is
$\sim 1.4 - 1.9 ~{\rm cm^{-2} s^{-1}}$.  Both the bounds on the flux and
that on the number of events depend on $E_0$ monotonically, getting
looser for larger $E_0$. This agrees with intuition: with the
decrease of $E_0$ the neutrino spectrum and the spectrum of the signal
of observed electrons fall more rapidly with energy. This implies that
the \df\ is constrained by the uncertainty on the number of events in
the lowest one or two energy bins (see figure \ref{malek_figure}).  In
contrast, with a less steep spectrum (larger $E_0$) part of the signal
could be in the higher energy bins, resulting in a looser constraint.

For the energy spectra used in the SK analysis \cite{Malek:2002ns},
e.g., $E_0=5.35$ MeV, we obtained a limit that is about 25\% looser
than the SK result, Eq. (\ref{sklim}): $\Phi_{\barnue}(E>19.3 ~{\rm
MeV})\simeq 1.49~{\rm cm^{-2} s^{-1}}$, compared to
$\Phi_{\barnue}(E>19.3 ~{\rm MeV})\simeq 1.2~{\rm cm^{-2} s^{-1}}$ of
ref. \cite{Malek:2002ns}.  We have checked that a 17\% discrepancy is
due to our using a more precise cross section for inverse beta decay,
while the remaining 8\% difference should probably be attributed to
details in the data analysis of the SK collaboration that we do not
have access to. It should be considered as an error associated to the specific method of analysis used.

The upper limit on the diffuse flux of $\barnue$ above 11.3 MeV varies
between $\sim 4.9 -13.4 ~{\rm cm^{-2} s^{-1}}$.  The allowed flux is
larger for smaller $E_0$, i.e., for steeper spectrum: indeed,
a spectrum that falls rapidly in energy allows a large flux below the
SK threshold, while still giving a sufficiently small number of events
in the SK energy window. 

Figure \ref{limits_barnue} completes the description of the $\barnue$
flux bounds, by showing how these vary with the energy interval, which
was taken of the form $[ E_{thr}, \infty]$.  This figure is of guidance
for experimental projects that may have different energy threshold
than SK.  Notice how around $ E_{thr} \sim 20$ MeV the flux limit has
little dependence on $E_0$ \footnote{We did not find a
value of $E_{thr}$ for which the dependence on $E_0$ is null.  The
curves in fig. \ref{limits_barnue}, as well as those for other
channels, figs. \ref{limits_nue}-\ref{limits_antinumu},  do \emph{not}
meet in one point, even though the figures might give such illusion
due to the use of the logarithmic scale.}.
This because $ E_{thr} \sim 20$ MeV corresponds to the interval of the
SK analysis, and for this reason it is the most strongly constrained.
Due to the exponential dependence of the \df\ on energy, a small
variation of the flux --  within the statistical errors of SK -- at
20-30 MeV corresponds to a large variation of the same flux at lower
or higher energy, and this explains the strong dependence of the
bounds on  $E_0$ for $ E_{thr}$  substantially  different from 20
MeV.

\subsection{Results for $\nue$}
\label{nueresults}

\begin{figure}[htbp]
  \centering
 \includegraphics[width=0.90\textwidth]{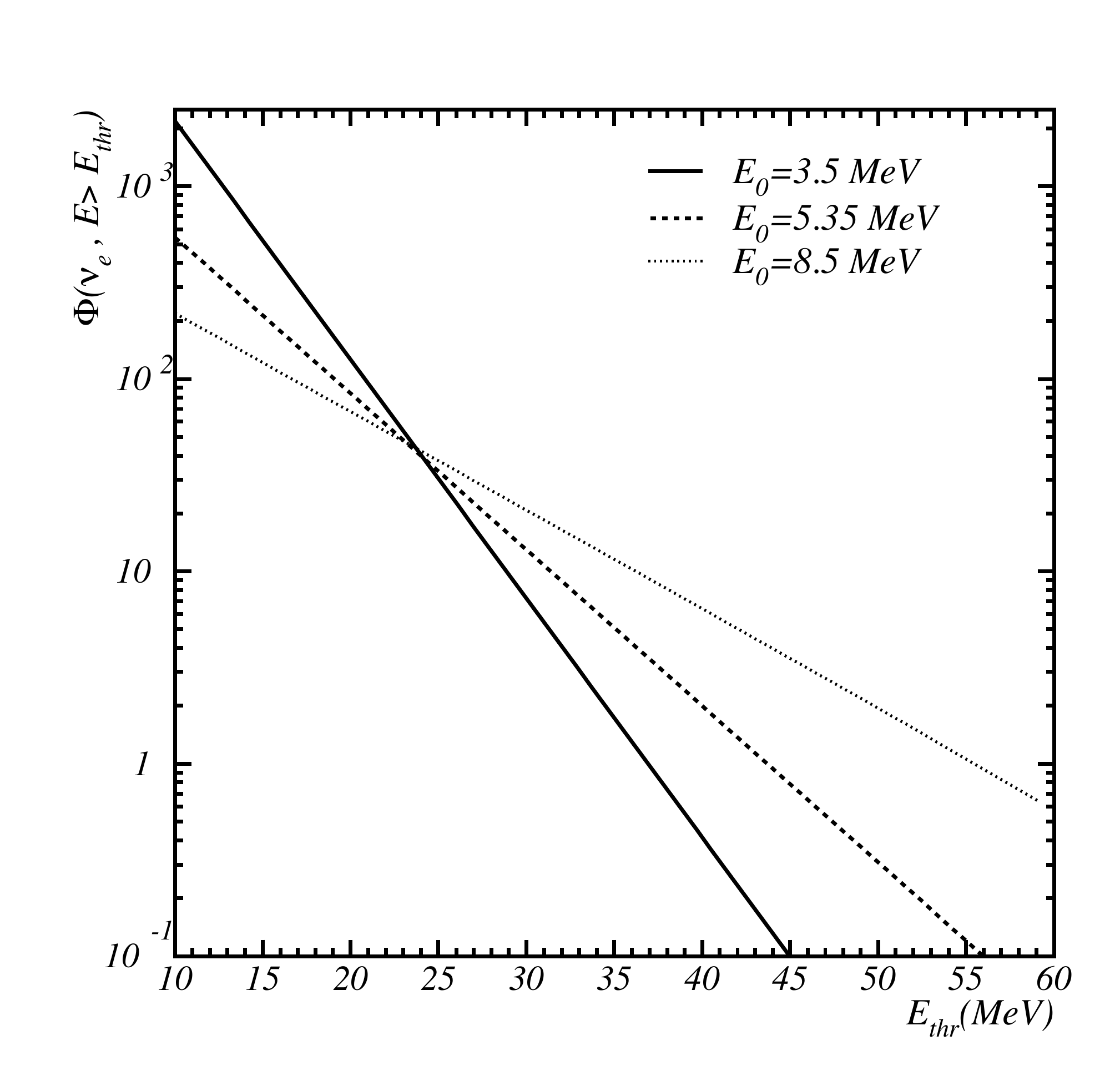}
\caption{The same as fig. \ref{limits_barnue} for  $\nue $.}
\label{limits_nue}
\end{figure}

The upper limits on the $\nue$ component of the \df, and on the
associated number of events, are given in Table \ref{tab_nue} and
fig. \ref{limits_nue}. In Table \ref{tab_nue} we present the limits
for the $\nue$ flux above 11.3 MeV and 19.3 MeV to allow direct
comparison with the $\barnue$ channel, where these thresholds are most
relevant. We also give results for the same energy window of the SNO
search,  $E = 22.9  -36.9$ MeV, also for the purpose of comparison.
 \begin{table*}
\centering
\begin{tabular}{| l || l || l | l | l |l}
\hline
\hline
$E_0/{\rm MeV}$  & $N_{90\%}$  & \multicolumn{3}{|c|}{limit on $\Phi_{\nue}$ (${\rm cm^{-2} s^{-1}}$)} \\ 
\hline
\hline
 &  & $E/{\rm MeV}>11.3$  & $22.9<E/{\rm MeV}<36.9$  & $E/{\rm MeV}>19.3$ \\
\hline
\hline
3.5 &   4.80 & $1.51 \cdot  10^3$ & 53.9  & $1.54 \cdot 10^2$   \\
\hline
3.8 &  5.17 & $1.16 \cdot 10^3$   & 53.4  &$ 1.41\cdot 10^2$    \\
\hline
5.35 &  8.86 & $4.27 \cdot 10^2$ & 45.3  &  95.8 \\
\hline
5.5 & 9.41 & $3.99 \cdot 10^2$ &  44.6  & 93.1 \\
\hline
6.5 &14.4  & $ 2.77 \cdot 10^2$ & 41.0   & 80.8  \\
\hline
7.5 &  23.5 & $2.20  \cdot 10^2$   & 39.6  & 75.7 \\
\hline 
8.5 &  33.3 & $1.88  \cdot 10^2$  & 38.8  & 73.3 \\
\hline
\hline
\end{tabular}
\caption{The same as Tab. \ref{tab_barnue} for the $\nue $ channel.}
\label{tab_nue}
\end{table*}

Similarly to the $\barnue$ channel, we see that larger $E_0$
corresponds to looser limit on the number of events, which can be as
high as $\sim 33$.  The upper bound on the flux of $\nue$  is
$1.5\cdot 10^3~ {\rm cm^{-2} s^{-1}}$ ($1.5 \cdot 10^2~ {\rm cm^{-2}
s^{-1}}$) above 11.3 MeV (19.3 MeV). It is about two orders of magnitude larger than the
 limit on the $\barnue$ flux in the same energy interval, due to the smaller
detection cross section.  
Besides the overall scale, the $\nue$ cross section (total of both
detection processes)  differs from that of inverse beta
decay in its rising faster with energy. This implies a less steep fall
of the spectrum of the observed leptons, and therefore an increased
possibility to have \df\ events distributed in the high energy bins
within the statistical errors. That is why the limits on the numbers
of events are generally looser for $\nue$.  The same difference in the
energy dependence of the cross section explains why in the $\nue$
channel the correlation between lower $E_0$ and larger flux allowed is
more pronounced.

In the interval $E = 22.9  -36.9$ MeV we find the $\nue$ flux to be
smaller than $54~ {\rm cm^{-2} s^{-1}}$. Such constraint improves on
the  SNO result in the same interval, Eq. (\ref{snolim}), and
therefore represents the best \emph{ direct} bound to date.

We refer to Fig. \ref{limits_nue} for the bounds in other energy
intervals; the qualitative features of the figure are similar to those
of fig. \ref{limits_barnue}.

\subsection{Results for $\nux$ and $\barnux$} \label{nuxresults}
\begin{figure}[htbp]
  \centering
 \includegraphics[width=0.90\textwidth]{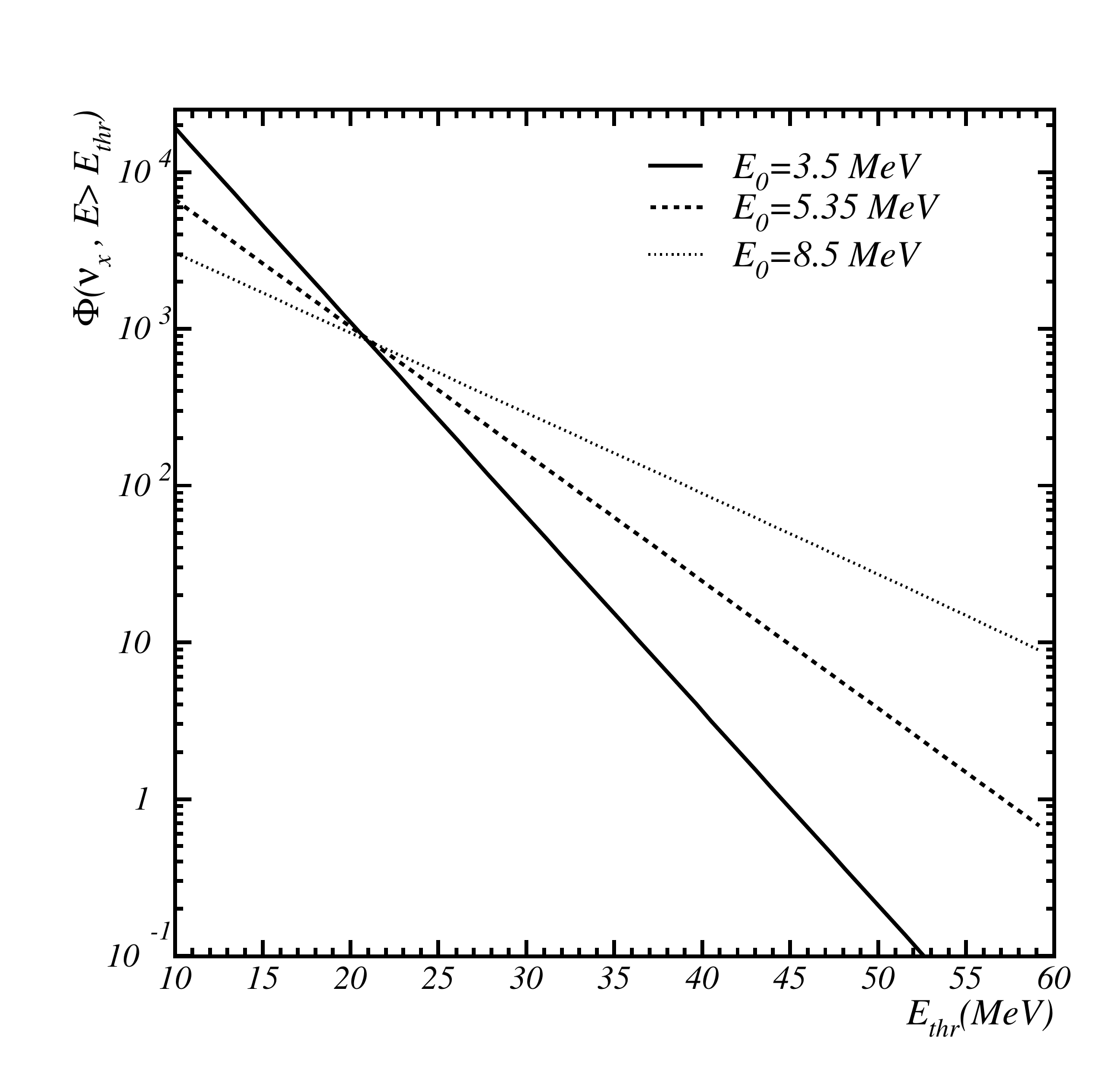}
\caption{The same as fig. \ref{limits_barnue} for  $\nux $. }
\label{limits_numu}
\end{figure}

\begin{figure}[htbp]
  \centering
 \includegraphics[width=0.90\textwidth]{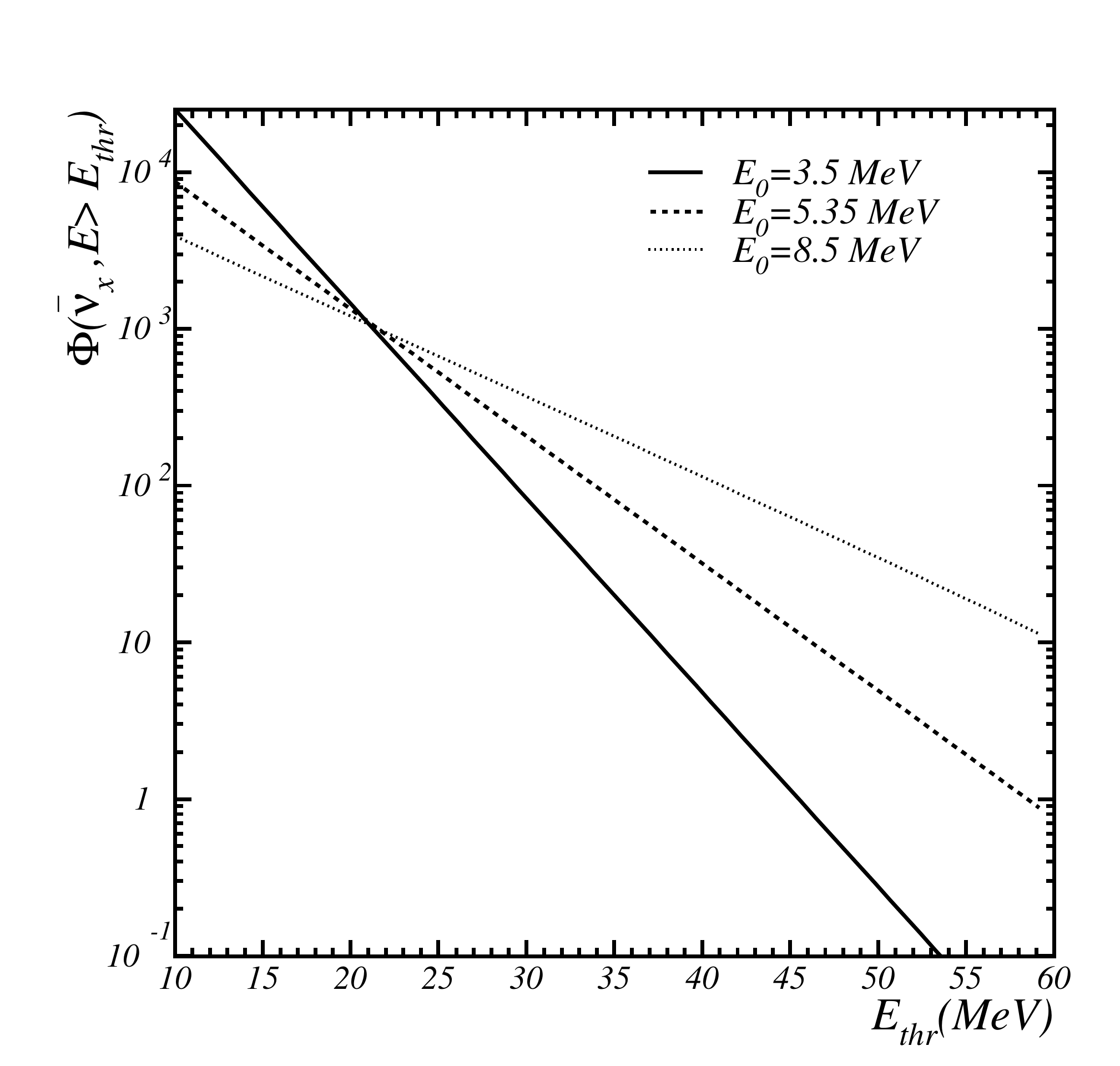}
\caption{The same as fig. \ref{limits_barnue} for  $\barnux $. }
\label{limits_antinumu}
\end{figure}

The upper bounds on $N_{90\%}$ and on the flux for $\nux$ and $\barnux$,
shown in Tables \ref{tab_nux} and \ref{tab_barnux}, and figures
\ref{limits_numu} and \ref{limits_antinumu}, are of the same order of
magnitude, with small differences reflecting differences in the cross
sections of neutrinos and antineutrinos.  For $E>19.3 $ MeV (Tables \ref{tab_nux} and \ref{tab_barnux}) the limit on the flux is
$\sim ( 1.0 - 1.35)\cdot 10^3 ~ {\rm cm^{-2} s^{-1}}$ for $\nux$ and
$\sim (1.3 - 1.8)\cdot 10^3 ~ {\rm cm^{-2} s^{-1}}$ for $\barnux$.
These improve by about four orders of magnitude on the previous best
limits from LSD, Eq. (\ref{lsdmutau}), which refer to the interval
$20-100$ MeV.

Due to the rough equipartition of energy among the six neutrino
species predicted by theory, it is reasonable to expect the $\nux$ and
$\barnux$ components of the \df\ to be comparable to the electron
flavor ones, that are more strongly constrained.  Therefore, the
direct limits we found here for $\nux$ and $\barnux$, while being the
best available,  are probably far from realistic values of these
fluxes.   They are nevertheless important for their observational,
model-independent character, and as a reference for future, more
sensitive, searches.

 \begin{table*}
\centering
\begin{tabular}{| l || l || l | l | }
\hline
\hline
$E_0/{\rm MeV}$  & $N_{90\%}$  & \multicolumn{2}{|c|}{limit on $\Phi_{\nux}$ (${\rm cm^{-2} s^{-1}}$)} \\ 
\hline
\hline
 &  & $E/{\rm MeV}>11.3$  &  $E/{\rm MeV}>19.3$ \\
\hline
\hline
3.5  & 4.38 &$1.32 \cdot 10^4$   &$1.35 \cdot 10^3$     \\
\hline
3.8 & 4.51  &$1.08 \cdot 10^4$   &$1.31 \cdot 10^3$     \\
\hline
5.35 & 5.24  &$5.29 \cdot 10^3$&$1.18 \cdot 10^3$    \\
\hline
5.5 & 5.32  &$4.99 \cdot 10^3$    &$1.17 \cdot 10^3$   \\
\hline
6.5 & 5.87  &$3.78 \cdot 10^3$ &$1.11 \cdot 10^3$   \\
\hline
7.5 &6.45  &$3.08 \cdot 10^3$   &$1.06 \cdot 10^3$ \\
\hline 
8.5 & 7.06 &$2.66 \cdot 10^3$   &$1.02 \cdot 10^3$      \\
\hline
\hline
\end{tabular}
\caption{The same as Tab. \ref{tab_barnue} for  $\nux $.}
\label{tab_nux}
\end{table*}

\begin{table*}
\centering
\begin{tabular}{| l || l || l | l | }
\hline
\hline
$E_0/{\rm MeV}$  & $N_{90\%}$  & \multicolumn{2}{|c|}{limit on $\Phi_{\barnux}$ (${\rm cm^{-2} s^{-1}}$)} \\ 
\hline
\hline
 &  & $E/{\rm MeV}>11.3$  &  $E/{\rm MeV}>19.3$ \\
\hline
\hline
3.5  &4.38  &$1.74 \cdot 10^4$   &$1.77 \cdot 10^3$     \\
\hline
3.8 & 4.51  &$1.41 \cdot 10^4$   &$1.72 \cdot 10^3$      \\
\hline
5.35 & 5.22  &$6.81 \cdot 10^3$  &$1.53 \cdot 10^3$   \\
\hline
5.5 &5.30  &$6.48 \cdot 10^3$     &$1.51 \cdot 10^3$    \\
\hline
6.5 &5.82  &$4.88 \cdot 10^3$   &$1.43 \cdot 10^3$   \\
\hline
7.5 & 6.40  &$3.96 \cdot 10^3$    &$1.36 \cdot 10^3$   \\
\hline 
8.5 & 7.00  &$3.35 \cdot 10^3$    &$1.31 \cdot 10^3$   \\
\hline
\hline
\end{tabular}
\caption{The same as Tab. \ref{tab_barnue} for  $\barnux $.}
\label{tab_barnux}
\end{table*}

\subsection{Interpretation of the results:  luminosity bounds} \label{interpretation}

What constraints can be drawn from our results on the \n\ emission from a supernova? 
A full address of this question requires a detailed analysis of all
the uncertainties, which we do not perform here.  However, as an
example we study what can be concluded on the \n\ luminosity if the
\sn\ rate is known, which is likely to be the case in the near future,
after data from SNAP \cite{snap} and JWST \cite{jwst} become
available. 

To extract the luminosity in a given flavor of an individual \n\ burst, one needs to go beyond the simple parametrization of the diffuse flux in Eq. (\ref{exp}).
We use the analytical description by Lunardini
\cite{Lunardini:2006pd}, which is valid within a few per cent above
the \sk\ threshold and is obtained from Eq. (\ref{flux}) with a power
law form of $R_{SN}$  and a \n\ spectrum at the exit of the star
(after oscillations) of the form of an exponential times a power law
\cite{Keil:2002in}.  It reads:
\be
\Phi \simeq  \frac{R_{SN}(0) L c}{H_0 \Gamma(2+\alpha) E^2_0} \left( \frac{E}{E_0} \right)^{\alpha-\eta -1} \left(\Gamma\left[\eta+1 ,\frac{E}{E_0} \right] -  \Gamma\left[ \eta+1 ,(1+z_{max}) \frac{E}{E_0} \right] \right)~,
\label{lum-anal}
\ee
where $L$ is the total energy (in the individual flavor) of the \n\
burst leaving the star, and  $\alpha \simeq 2 -5 $ is a parameter
describing its spectral shape \cite{Lunardini:2006pd}.  The average
energy of the \ns\ leaving the star is $\langle E \rangle \simeq E_0
(1+\alpha)$.  We have $\eta=\alpha + \beta - 3 \Omega_m/2$, where
$\beta$ describes the cosmological supernova rate: $R_{SN}\simeq
(1+z)^\beta$.  $z_{max} \sim 1 $ is a break in the redshift
distribution of \sne.   We fix $R_{SN}$ to be the best piecewise fit
to the data found by Beacom and Hopkins \cite{Hopkins:2006bw}:
$R_{SN}\simeq 2.0 \cdot 10^{-4}~{\rm yr^{-1} Mpc^{-3}}$, $\beta=3.28$
and $z_{max}=1.0$.  While Eqs. (\ref{exp}) and  (\ref{lum-anal})
differ only minimally in the energy dependence
\cite{Lunardini:2006pd}, the second is more adequate for our purpose
because in it the dependence on the different parameters is explicit. 

For every value of $E_0$ and $\alpha$, the combination of
Eq. (\ref{lum-anal}) and of our flux limits (Tables
\ref{tab_barnue}-\ref{tab_barnux}) gives an upper limit on the
integrated luminosity $L$, which can then be compared to the
theoretical expectation of $L\sim 5 \cdot 10^{52}~{\rm ergs}$ per \n\
flavor.

\begin{figure}[htbp]
  \centering
 \includegraphics[width=0.43\textwidth]{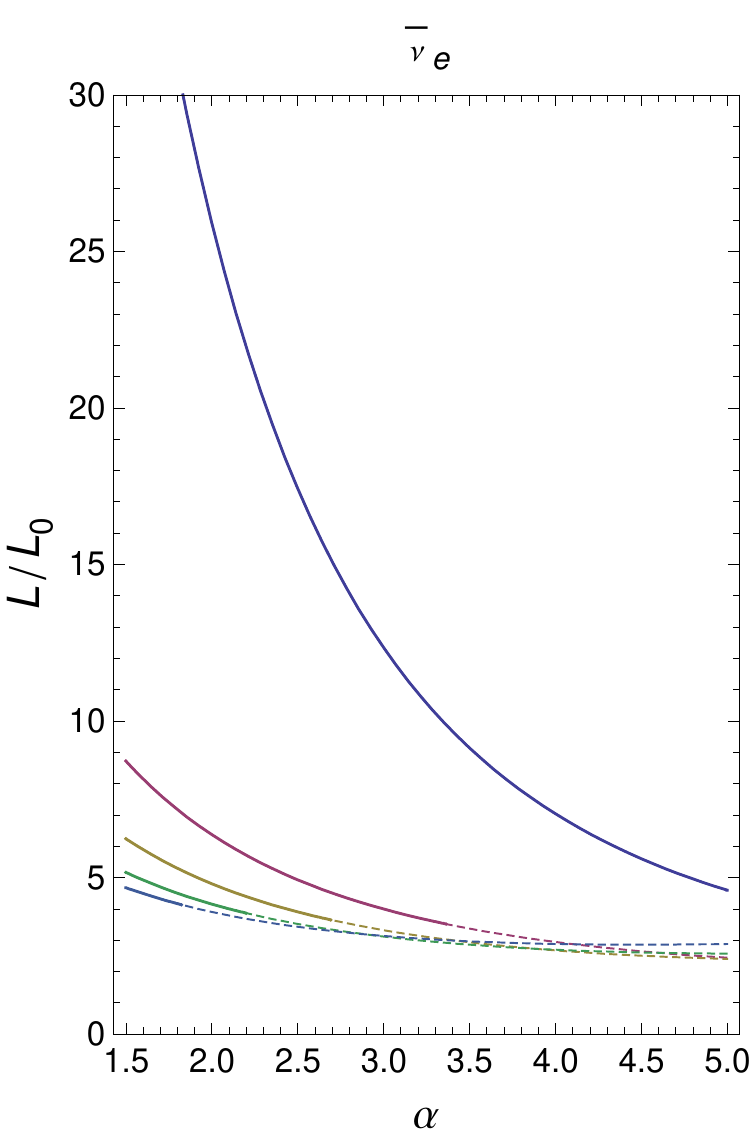}
 \includegraphics[width=0.45\textwidth]{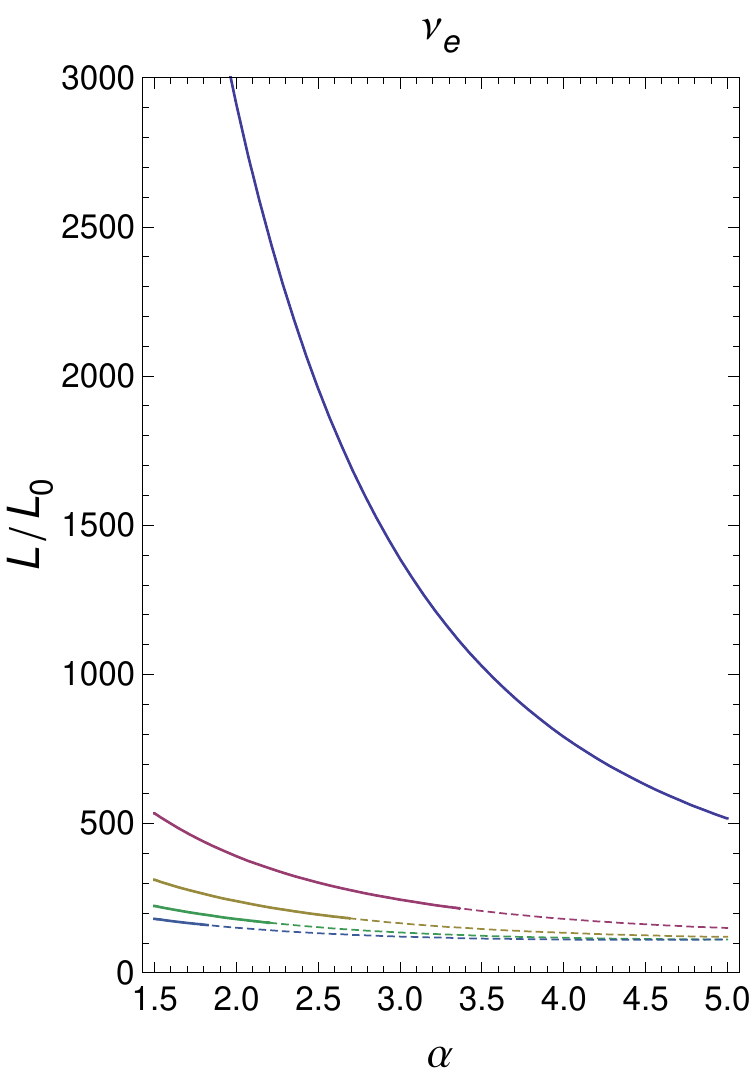}
 \includegraphics[width=0.45\textwidth]{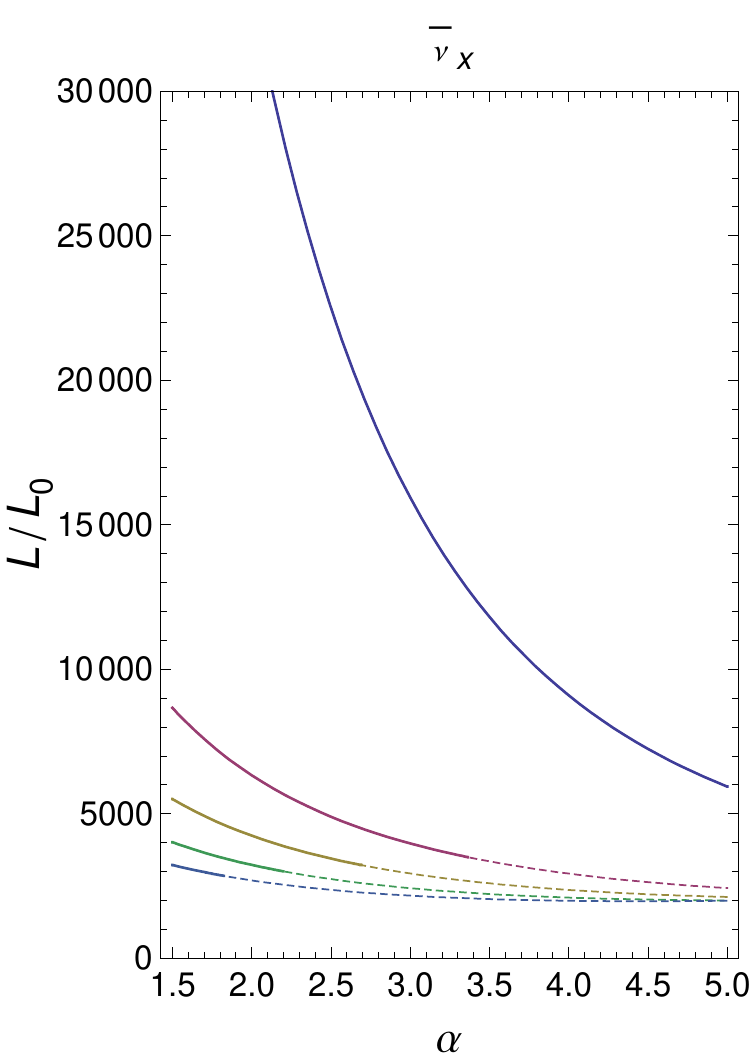}
 \includegraphics[width=0.45\textwidth]{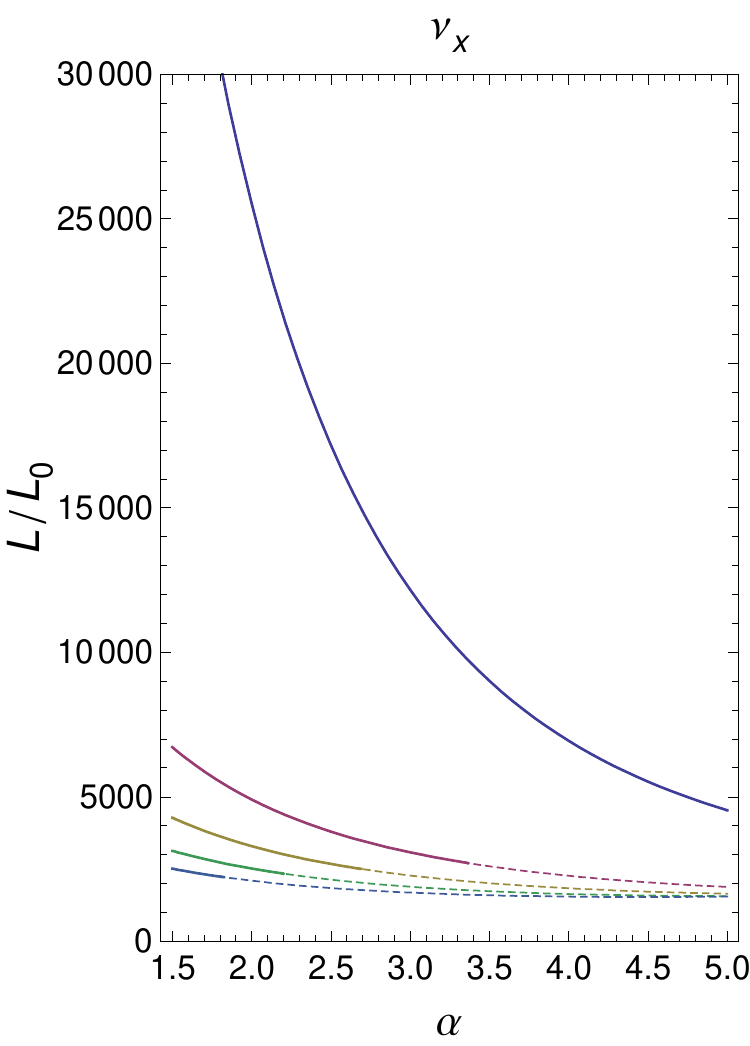}
\caption{90\% C.L. upper limits on the neutrino luminosity (in units
of $L_0=10^{52}$ ergs) in the different species at the exit of the
star, as a function of the ``shape'' parameter $\alpha$.  The curves,
from upper to lower, correspond to $E_0/{\rm MeV}=3.5,5.5,6.5,7.5,8.5$ in every
panel.  The solid curves were obtained by imposing a naturalness
condition on the average energy of the \n\ spectrum at the exit of the
star: $9<\langle E \rangle/{\rm MeV} < 24$. Notice the large
difference in the vertical scales between the different panels.}
\label{lum}
\end{figure}

Results are shown in fig. \ref{lum}, where we have imposed $9<\langle
E \rangle/{\rm MeV} < 24$ as naturalness constraint (solid curves). We
see that, for $E_0 \gta 5.5$~MeV, the upper limit on the $\barnue$
luminosity approaches, or even crosses, the theoretically interesting
range, an intriguing fact that has been already observed in \cite{Yuksel:2005ae}.
Softer \n\ spectra, however, allow \n\ the \n\ luminosity to be
several times larger than the theoretical central value.   This is
expected, since softer \n\ spectrum means smaller fraction of the
total energy above the \sk\ threshold.  A further loosening of the
luminosity constraint is expected if  the uncertainty on the \sn\ rate
is included \cite{Lunardini:2005jf}. 

The luminosity bounds for the other \n\ species are far from the
theoretical range, reflecting the weaker flux limits.  They could be
useful to constrain the presence hidden sources that could add to the
expected \sn\ \n\ flux. \ceci{cite people?  remove this statement?}

\section{Summary and discussion}
\label{discussion}

We have analyzed the data of the search for the diffuse supernova
neutrino flux published by the SK collaboration in 2003
\cite{Malek:2002ns}.   Our work extends the analysis of collaboration
itself in a number of ways. 
First, it considers all the relevant channels,  one  at a time (i.e.,
with all the signal attributed to a certain channel), resulting in
bounds on the fluxes of all the neutrino species. 
Second, it takes into account a wider range of neutrino spectra,
including the softer ones that have been suggested in the literature
recently (Table \ref{tableslopes}).   It also uses a more precise
cross section for inverse beta decay, that rises more slowly with
energy and therefore implies a looser bound on the $\barnue$ component
of the flux.    Results have been given for several different energy
intervals motivated by near future experiments.

A first result is that the flux of $\barnue$ is bound, at 90\% CL, to be smaller than $1.4-1.9~{\rm
cm^{-2} s^{-1}}$ for $E>19.3$ MeV and than $4.9-13.4~{\rm
cm^{-2} s^{-1}}$ above a 11.3 MeV threshold.  The flux intervals
account for the variation with the neutrino energy spectrum, and the
discrepancy with the SK limit, Eq. (\ref{sklim}), is explained by the
different cross section used, up to a minor 8\% difference.

The limits in the $\nue$ channel improve on the previous limit from
SNO, Eq. (\ref{snolim}): at 90\% C.L. the $\nue$ flux must be smaller
than $54~{\rm cm^{-2} s^{-1}}$ in the interval $22.9-36.9$ MeV even
for the softest neutrino spectrum, which represents the most
optimistic case.  The flux is allowed to be as large as   ${\mathcal O}(10^2)~{\rm
cm^{-2} s^{-1}}$ (${\mathcal O}(10^3)~{\rm
cm^{-2} s^{-1}}$) above 19.3 (11.3) MeV energy. 

The upper limits on the non-electron species are in the range of ${\mathcal O}(10^3)~{\rm
cm^{-2} s^{-1}}$ to ${\mathcal O}(10^4)~{\rm
cm^{-2} s^{-1}}$ for the same two thresholds, and improve by several
orders of magnitude over the previous bounds from LSD,
Eq. (\ref{lsdmutau}).

Our results represents the most one can get from this generation of
detectors on the diffuse flux, up to minor improvements that might
come from increased statistics at SuperKamiokande (see the preliminary
results in \cite{skiida}).  Considering that theoretical predictions
of the \df\ in each neutrino species range from touching the current
SK limit for $\barnue$ down to values 20 times smaller
(Tab. \ref{tableslopes}), it is clear that SK might not be able to see
the \df\ in the $\barnue$ channel and almost certainly will never
detect it in other channels. This strongly reinforces the already strong case for new detectors with enhanced sensitivity to the \df.  

 \begin{table*}
\centering
\begin{tabular}{| l |l  |l  |l  |l |}
\hline
\hline
 concept  & channel & name of project & energy &  SK flux limit \\ 
   & &  &(MeV)  & (${\rm
cm^{-2} s^{-1}}$) \\ 
\hline
\hline
liquid  & $\barnue$ & LENA  \cite{Wurm:2007cy}  & 9.5-30  &    $5.5-22.3$  \\
 scintillator  &  & (at Phy\"asalmi)   & &   \\
 \hline
liquid  & $\barnue$ & LENA  \cite{Wurm:2007cy}  & 8.2-27.2 &   $6.3-32$  \\
 scintillator  &  & (at Wellington)   & &   \\
 \hline
 water   & $\barnue$ &HyperKamiokande, UNO,  & 19.3-80  &  $1.4-1.9$ \\
  & & MEMPHYS (generic) \cite{Jung:1999jq,Nakamura:2003hk,Mosca:2005mi}  & &  \\
 \hline
 water+Gd   & $\barnue$& GADZOOKS   &15-30  &   $2.6 -4.6$ \\
  & &(at Kamioka) \cite{Autiero:2007zj} & &  \\
\hline
 water + Gd   & $\barnue$&  GADZOOKS (generic)  \cite{Beacom:2003nk}  &11.3-80  & $5-13$ \\
\hline
 liquid argon  & $\nue$&  GLACIER, LANDD  \cite{Cocco:2004ac,Ereditato:2005yx,Cline:2006st}&  16-40   & $(3.9-8.2)\cdot 10^2$ \\
\hline
\hline
\end{tabular}
\caption{  Flux limits from SK at 90\% C.L. for the energy intervals
and channels that are relevant to the four most discussed  future \df\
detectors (see text for details).   We distinguish between generic
designs and detailed studies that include estimates of the background
at specific sites (see Table 8 in \cite{Autiero:2007zj}). The
intervals in the flux correspond to varying the  neutrino spectrum in
the range shown in Table \ref{tableslopes}.}
\label{tabfinal}
\end{table*}

With this in mind, we have given constraints on the flux in different
energy intervals that could be relevant for  future searches.  They
represent the minimum sensitivity that these searches should have to
improve on the current status. To summarize, and also to add detail,
in Table \ref{tabfinal} we give the SK bound on the \df\ for the
specific setup (neutrino species and energy interval) of  the four
different concepts  that are most discussed for the future. These are
a Megaton water Cerenkov detector
\cite{Jung:1999jq,Nakamura:2003hk,Mosca:2005mi}, a 50 kt liquid
scintillator experiment \cite{MarrodanUndagoitia:2006re,Wurm:2007cy},
a liquid Argon chamber \cite{Ereditato:2005yx,Cline:2006st} and the SK
tank with Gadolinium trichloride dissolved in it \cite{Beacom:2003nk}.
For some projects, we consider both generic and site-specific designs,
for which the background, and therefore the energy window of
sensitivity, have been calculated in detail\footnote{The energy
intervals for the site-specific designs are nevertheless tentative,
because they depend on the luminosity and spectrum of the \df.  Rather
than optimal intervals, they should be considered as likely choices
that an experimental collaboration would adopt for a data analysis.}. 
From the  Table one can immediately see how crucial it is to have a
detection threshold as low as possible, since the \df\ falls
exponentially with energy. 

The SK bounds in Table \ref{tabfinal} will certainly be a practical
benchmark for the planning of future detectors, as they allow to give
a quantitative estimate of how much a given design will improve over
SK.  Moreover, our bounds will be an important component of a global
analysis of multiple data sets  once new data are available from
different experiments. 

We conclude with the comment that the detection of the diffuse flux
of supernova neutrinos is an opportunity that the underground science
community can not afford to miss, because of its implications for
fundamental physics and astrophysics/cosmology.  Since the flux is
continuous in time, the barrier to its observation is only
technological. Once the barrier is overcome, the experimental study of
supernova neutrinos will transition from the realm of rare event to
that of regular, ongoing data taking and analysis.  Our results
represent the status of the art of current searches and the ultimate
sensitivity -- up to minor improvements -- of the existing detectors.
It marks where  the present generation of experiments leaves off and
the new one will take on, in the next decade or so, to bring this
field to maturity.

\subsection*{Acknowledgments}
C.L. acknowledges support from the ORNL grant of the Institute of Nuclear
Theory (INT) of Seattle, where this work was initiated,  from Arizona
State University, and from the RIKEN BNL Research Center (RBRC).
O.L.G. Peres  acknowledges support from FAPESP, CNPq and FAEPEX; he is
grateful to the INT for hospitality. We are especially indebted to
M.S. Malek and R.J. Wilkes for useful exchanges.



\providecommand{\href}[2]{#2}\begingroup\raggedright\endgroup

\end{document}